# A computational model for synaptic message transmission

Lizhi Xin[1], Kevin Xin[2], Houwen Xin[3, *]

[1]Hefei, P. R. China
[2]Chicago, USA
[3]Department of Chemical physics USTC, Hefei, Anhui, P. R. China
[*]hxin@ustc.edu.cn

## Abstract

A computational model incorporating insights from quantum theory is proposed to describe and explain synaptic message transmission. We propose that together, neurotransmitters and their corresponding receptors, function as a physical "quantum decision tree" to "decide" whether to excite or inhibit the synapse. When a neurotransmitter binds to its corresponding receptor, it is the equivalent of randomly choosing different "strategies"; a "strategy" has two actions to take: excite or inhibit the synapse with a certain probability. The genetic programming can be applied for learning the observed data sequence to simulate the synaptic message transmission.

## Introduction

Our self-consciousness arises entirely from the results of activities in the nervous system [1]. And these activities are the transfer of messages among the synapses between neurons. Neurons transmit information through a synapse with the help of neurotransmitters that are stored in vesicles at the end of its axons. Then there are action potentials that propagate along those axons to release neurotransmitters, which are then met by receptors on the postsynaptic membrane responding to certain, specific neurotransmitters, forming ion channels with switching function to excite or inhibit postsynaptic neurons. When the ion channel is opened, the incoming ions will reduce the synaptic potential difference inside and outside the cell membrane, which is called depolarization, resulting in excitation; the outgoing ions will increase the potential difference, which is called hyperpolarization, resulting in the inhibition of neurons. While the ion channel is opened or closed, the synaptic potential changes continuously to form excitation or inhibitory synaptic inputs into the neuron body for integration, when integrated synaptic potential exceeds the threshold, the all-or-nothing action potential will be triggered.

We will propose a computational model based on our quantum decision theory for synaptic message transmission. In our quantum decision theory [2], quantum decision tree (qDT) is the core concept: a qDT consists of several strategies from which the decision-maker randomly chooses a strategy to guide their action (with certain subjective beliefs). In this paper, we further postulate that neurotransmitters and their corresponding receptors together function as a physical qDT: a neurotransmitter binding to different receptors is equivalent to randomly choosing a "strategy" (excite or inhibit the synapse with a certain probability). We believe that the computational models of neurotransmitters and receptors as qDTs are constructed by the brain over many years of evolution by constantly learning and adapting to the environment.

## A computational model for neurotransmission

A synapse has two states: excitation or inhibition; whether the state of the synapse is either excitation or inhibition, which all depends on all neurotransmitters binding to corresponding receptors. This uncertain state can be represented by a superposition state in (1):

$$|\psi\rangle = c_1|q_1\rangle + c_2|q_2\rangle \qquad |c_1|^2 + |c_2|^2 = 1 \qquad (1)$$

Where $|q_1\rangle$ denotes a state in which the synapse is excited and $|q_2\rangle$ denotes a state in which the synapse is inhibited; $|c_1|^2$ is the probability of the excitation, $|c_2|^2$ is the probability of the inhibition.

When a neurotransmitter binds to a corresponding receptor it either excites or inhibits the synapse; the "outcome" of the neurotransmitter-receptor binding event is uncertain (excited or inhibited) because it depends on which receptor the neurotransmitter binds to; the uncertainty of the neurotransmitter-receptor binding event can be represented by a superposition in (2):

$$|\phi\rangle = \mu_1|a_1\rangle + \mu_2|a_2\rangle \quad |\mu_1|^2 + |\mu_2|^2 = 1 \tag{2}$$

Where $|a_1\rangle$ denotes the neurotransmitter-receptor binding to excite the synapse, and $|a_2\rangle$ denotes the neurotransmitter-receptor binding to inhibit the synapse. $|\mu_1|^2$ is the neurotransmitter-receptor binding event's probability of exciting the synapse; $|\mu_2|^2$ is neurotransmitter-receptor binding event's probability of inhibiting the synapse.

Before a neurotransmitter binds to a receptor, the state of the neurotransmitter-receptor binding is in a pure state ($\rho = |\phi\rangle\langle\phi|$): a superposed state in which it can both excite and inhibit the synapse at the same time. But in reality, a neurotransmitter-receptor binding can't excite and inhibit the synapse simultaneously. This pure state is when the states of excite and inhibit are superposed in the neurotransmitter-receptor binding. Then when the neurotransmitter binds to a corresponding receptor, the state of the neurotransmitter-receptor binding event is then transformed from that pure state into a mixed state $\rho'$, which is when the neurotransmitter-receptor binding either excites or inhibits the synapse. Basically, this transformation is when the neurotransmitter-receptor binding event "chooses" from one of the available actions, with action $a_1$ being excite the synapse with probability $p_1$ and action $a_2$ being inhibit the synapse with probability $p_2$ in (3)

$$NT \xrightarrow{\text{binds}} R_i: \rho = |\phi\rangle\langle\phi| \xrightarrow{\text{"choose"}} \rho' = p_1|a_1\rangle\langle a_1| + p_2|a_2\rangle\langle a_2| \tag{3}$$

Where NT denotes neurotransmitter and $R_i$ denotes a corresponding receptor; the density operator of a pure state ($\rho = |\phi\rangle\langle\phi|$) is a 2X2 matrix that can be constructed by 8 basic quantum gates as well as three operators, in effect constructing a quantum decision tree (qDT).

Postulate 1: a neurotransmitter and receptor work together to function as a qDT.

Postulate 2: a neurotransmitter binding to a corresponding receptor is equivalent to selecting one strategy and take one action with certain probability, which is either excite or inhibit the synapse.

The qDT (4a) of Figure 1 consists of quantum gates D, T, and Z and has two strategies (4b, 4c):

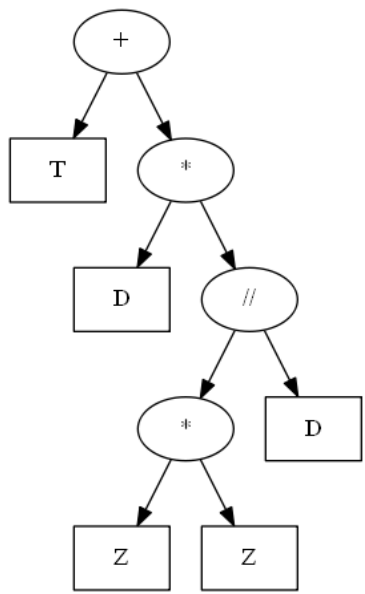

**Figure 1 A simulated qDT**

$$qDT = \left(T + \left(D * ((Z * Z)//D)\right)\right) \tag{4a}$$

- $S_1 = \left(T + (D * (Z * Z))\right) \rightarrow \hat{\rho} = |a_1\rangle\langle a_1|, p_1 = 1, p_2 = 0$ (4b)

- $S_2 = \left(T + (D * D)\right) \rightarrow \hat{\rho} = |a_2\rangle\langle a_2|, p_1 = 0, p_2 = 1$ (4c)

For the neurotransmitter-receptor binding event, this qDT provides two "strategies" $\{S_1, S_2\}$, and the neurotransmitter-receptor binding can randomly "choose" a "strategy" from the two and "apply" the chosen "strategy" that'll guide the neurotransmitter-receptor binding in "choosing" which "action" (to excite or to inhibit) to "take" (with a certain probability). If strategy $S_1$ is "chosen", then the neurotransmitter-receptor binding excites the synapse; if strategy $S_2$ is "chosen", the neurotransmitter-receptor binding inhibits the synapse.

Taking glutamate for example, there are two types of receptors: ionotropic and metabotropic receptors [3-4]. Studies have shown that when glutamate binds to ionotropic receptors, ion channels are formed, which excites synapses [5]; binding of glutamate to metabotropic receptors does not form ion channels, which inhibits synapses[6]. Our qDT model (neurotransmitter-receptor binding model) can be explained as: glutamate and receptors together "form" a gDT (4a), strategy $S_1$ is the glutamate binding to an ionotropic receptor (4b) to excite the synapse, and strategy $S_2$ is the glutamate binding to a metabotropic receptor (4c) to inhibit the synapse.

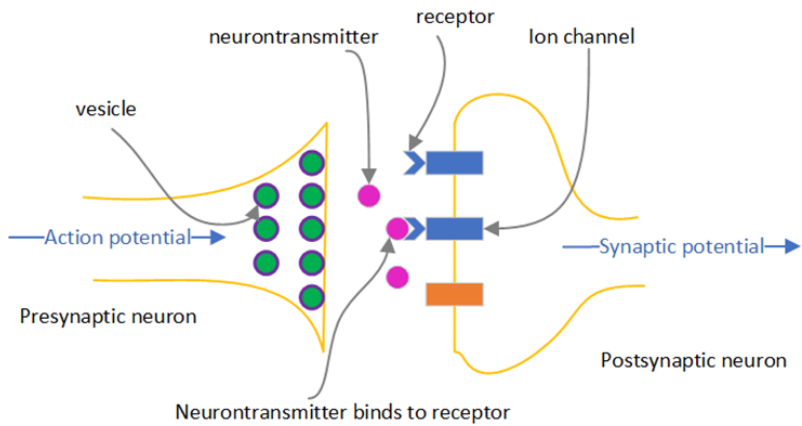

**Figure 2 Structure of neurotransmitter-receptor binding**

The two states of $q_1$ and $q_2$ signify that the synapse is exciting or inhibiting, respectively. Then the neurotransmitter-receptor have the two possible binding "actions", $a_1$ and $a_2$ being excite and inhibit, respectively. The function state of the synapse influences the neurotransmitter-receptor binding event; while simultaneously, all the N amount of times the neurotransmitter-receptor binding at the synapse determines the function state of the synapse (excite or inhibit) as shown in Figure 2. This interaction between the two, the function state of the synapse and the binding state of the neurotransmitter-receptor is what causes both the result of the binding and the state of the synapse to be uncertain. When a neurotransmitter binds to a receptor, there are four possible outcomes as shown in Table 1. For example, if the state of the synapse is excited and the neurotransmitter binds to the receptor to excite the synapse, then the expected value is $p_1$("reward"); if the state of the synapse is excited and the neurotransmitter binds to the receptor to inhibit the synapse, then the expected value is $-p_2$("punishment").

| State of synapse \ Neurotransmitter-receptor binding | $q_1$ | $q_2$ |
|---|---|---|
| $a_1$ | $p_1$ | $-p_1$ |
| $a_2$ | $-p_2$ | $p_2$ |

**Table 1** Possible outcomes of neurotransmitter-receptor binding.

The expected value of a kth neurotransmitter-receptor binding is in (5):

$$EV_k = \begin{cases} = p_1, \text{a glutamate binds to an ionotropic receptor to excite the synapse and the synapse is in a state of excitation} \\ = -p_1, \text{a glutamate binds to an ionotropic receptor to excite the synapse but the synapse is in a state of inhibition} \\ = -p_2, \text{a glutamate binds to a metabotropic receptor to inhibit the synapse but the synapse is in a state of excitation} \\ = p_2, \text{a glutamate binds to a metabotropic receptor to inhibit the synapse and the synapse is in a state of inhibition} \end{cases} \quad (5)$$

It is critical to note here that from the above four possible outcomes only one can actually happen. Because the state of a synapse can only be excited or inhibited and a neurotransmitter-receptor binding are only to excite or inhibit the synapse, in which only one of those combinations can occur in reality.

The neurotransmitter-receptor qDT model can be simulated by machine learning on historical data by genetic programming based on maximizing expected value of neurotransmitter-receptor binding event, and we believe that different neurotransmitters and their corresponding receptors are the result of years of evolution of the brain adapting to nature. The fitness function of genetic programming [7-8] represents the sum of all the $EV_k$ by machine-learning all the historical data as shown in (6):

$$f_{fitness} = \sum_{k=0}^{N} EV_k \quad (6)$$

The historical data of the state of the synapse can be represented as data sequences $\{(q_k, x_k)\} \; k = 1, \cdots, N$ for machine learning, where $q_k$ denotes the state of the synapse and $x_k$ denotes the synaptic potential. The evolutionary algorithm can study the data sequence $(q_k, x_k)$ through genetic programming to evolve an optimized qDT (neurotransmitter-receptor binding model) to simulate synaptic message transmission.

## Discussion

Throughout the years there has been great strides in the cognitive sciences, neurosciences, and biology, but we still don't understand how the single notes of neurons come together to form a symphony of communication that gives rise to mystical yet essential ways that all living beings behave and make decisions, basically what can be termed consciousness [9-13]. Ever since Descartes proposed the mind-body dualism, the trek of figuring out the ultimate question of our very existence officially began. And now about 300 years later, we still have not gotten much further than Descartes did. Though the advancements in science have given us many new insights, the evoking of modern physics, especially quantum mechanics has inspired some new thought-provoking theories [14-20], but even so we have only barely scratched the surface.

Ludwig Boltzmann once said, "The intimate connection of the mental with the physical is in the end given to us by experience." And we believe that experience can be represented or stored as data sequences at the synapse level. We postulate that it is these data sequences of 0s and 1s (excite or inhibit the synapse) that are the result of the neurotransmitter-receptor binding that eventually produces consciousness.

We postulate the neurotransmitter-receptor binding as a qDT, it holds because a qDT ultimately decides a yes or no choice by picking a strategy with two actions of different probability and choosing between the two. A neurotransmitter, when binding to a corresponding receptor has the same yes/no "decision" to make as well: excite or inhibit the synapse. Thus, it has a group of "strategies" available to be selected from just like the qDT: one group maybe to strongly excite the synapse, another maybe to strongly inhibit the synapse, another could be 50/50, and more groups of strategies may be available to be selected from. And each strategy only has those 2 potential actions: excite or inhibit. Just like how neurotransmitters themselves are the product of natural selection and have evolved throughout millions of years, the way it currently "decides" to "pick" what action to take (excite or inhibit) has too as well. Because the way a neurotransmitter-receptor binding event "chooses" whether to excite or inhibit the synapse is based on the current information obtained from the external environment (nature), and compares it with the past experience it gained by the surviving "fittest" ones that is stored in the cortex (memory), or as some would call it, an illusion.

This process of "reconstructing" an "illusion" of reality starts off with receiving information from the outside world with our bodies. Our bodies are full of sensors where we take in things from our surroundings. Our ears, skin, eyes, mouth, nose, hands, and practically every part of our body acts as sensors, sensors for vision, hearing, touch, taste, and smell that gives us all our senses. Once our bodies take in and receive that info, triggering off an action potential that kicks the brain into action, beginning to encode the external world's info. This first produces the initial individual neuron's 0/1 sequence at the synapse level, and then all the neurons' sequences come together to form another 0/1 sequence of an entire neural network, which is the "reconstruct" of how we eventually perceive reality. This neural network of all the individual neuron sequences combined together creates another sequence of 0s and 1s which is the eventual "illusion" of what we think reality is. Our brain reconstructs outside reality as our self-conscious, where we get our self-awareness, and it is different for every single one of us.

We believe that the information of the objective world is encoded in action potentials, and probabilistically transformed into subjective signals at the synapse level. Our proposed synapse's neurotransmission hypothesis states that the combination of neurotransmitter and receptor functions as the "quantum decision tree", which controls the exciting and inhibiting the synapse by deciding the firing rate of neurons. The firing pattern of the neural network expresses the decision of the brain, that is, subjective beliefs are converted from objective information, an attempt at bringing together the fathomless abyss.

Current research is purely theoretical, further research will apply our hypothesis to observed synapse data of various neurotransmitters and receptors (dopamine, serotonin, glutamate, etc.) to simulate synaptic message transmission for different neurotransmitter-receptor binding.

## Data availability
No datasets were generated or analyzed during the current study.

## Author contributions statement

## Additional information